\documentclass[aps,prd,twocolumn,showpacs,amsmath,amssymb,nofootinbib]{revtex4-1}
\usepackage{graphicx}
\usepackage{color}
\usepackage{times}
\usepackage{inputenc}
\usepackage{bm}
\usepackage{ulem}
\usepackage{multirow}
\usepackage{float}
\usepackage{url}
\usepackage{natbib}
\usepackage[colorlinks=true,citecolor=blue,urlcolor=blue,linkcolor=red]{hyperref}
\begin{document}
	%%%%
	\title{Dark matter admixed neutron star as a possible compact component
           in the GW190814 merger event}
	%%%%
	\author{H. C. Das$^{1,2}$}
	\email{harish.d@iopb.res.in}
	\author{Ankit Kumar$^{1,2}$}
	%\email{ankit.k@iopb.res.in}
	\author{S. K. Patra$^{1,2}$} 
	%\email{patra@iopb.res.in}
	%%%%
	\affiliation{\it $^{1}$Institute of Physics, Sachivalaya Marg, Bhubaneswar 751005, India}
	\affiliation{\it $^{2}$Homi Bhabha National Institute, Training School Complex,	Anushakti Nagar, Mumbai 400094, India}
	%%%%
	\date{\today}
	\begin{abstract}
	We put constraints on the secondary component of GW190814 by analyzing the observational data of the event. The relativistic mean-field models are used to calculate the mass-radius profile and tidal deformability of the compact object, considering it as a massive neutron star with the presence of dark matter particles inside it. With the increase of dark matter percentage, the maximum mass, radius, and tidal deformability of the neutron star decreases. We observe that the predicted properties are well consistent with GW190814 observational data, suggesting the possibility of a dark matter admixed neutron star if the underlying nuclear equation of state is sufficiently stiff.
	\end{abstract}
	%%%%
	\maketitle
	%%%%
	\section{Introduction}
	The LIGO and Virgo collaboration detected the most enigmatic compact binary coalescence event (GW190814) involving a  black hole and a compact object with a mass range of 22.2--24.3 and  2.50--2.67 $M_\odot$ respectively \cite{RAbbott_2020}. Since the electromagnetic counterpart has not been found and has no measurable signature of tidal deformations from the gravitational wave (GW), the secondary component might be the lightest black hole or the heaviest neutron star (NS). Therefore, it is a big challenge to explain the nature of such compact object. There is a lot of debate to understand the mystery of the secondary component of the GW190814 event \cite{Most_2020, Vattis_2020, Tews_2020, ZhangAAS_2020, Lim_2020, Godzieba_2020, Huang_2020, Tan_2020, Fattoyev_2020, Roupas_2021, Biswas_2021}. The analysis of GW190814 data implies the possible nature of such a compact object as a NS only when (i) the equation of state (EOS) is very stiff \cite{Huang_2020} or (ii) it is a rapidly rotating compact object below the mass shedding frequency \cite{ZhangAAS_2020, Biswas_2021}. Thus, one can consider either of the two facts, which can produce the mass of the secondary component of GW190814 around 2.50 $M_\odot$.
	
	The binary NS merger event GW170817 \cite{Abbott_2017} has evoked to put an upper bound on the maximum mass of nonrotating NS. By combining the total binary mass of GW170817 inferred from GW signal with electromagnetic observations, an upper limit of $M_{{\rm max}}\leq2.17 \ M_\odot$ is predicted by Margalit {\it et al.} \cite{Margalit_2017}. Rezzolla {\it et al.} have put the upper bound by combining the GW observations and quasiuniversal relations as $M_{{\rm max}}\leq 2.16_{-0.15}^{+0.17}\ M_\odot$ \cite{Rezzolla_2018}. Further analysis by employing both energy and momentum conservation laws and the numerical-relativity simulations show that the maximum mass of a cold NS is bound to be less than $2.3\ M_\odot$ \cite{Shibata_2019}. Also, various massive pulsar discoveries constrained the EOSs of the supra-nuclear matter inside the core of the NS \cite{Demorest_2010, Antoniadis_2013, Cromartie_2019}. These observational data also suggest strong constraints on the maximum mass of the slowly rotating NS with a lower bound of $\sim 2\ M_\odot$, which discarded many EOSs. Recently, the Neutron Star Interior Composition Explorer (NICER) data also imposed stringent constraints on the mass and radius of a canonical star from the analysis of PSR J0030+0451 data \cite{Miller_2019, Riley_2019, Raaijmakers_2019}. However, one cannot exclude the existence of a super-massive NS as the secondary component of GW190814. 
	
	Many energy density functionals have been formulated to study the finite nuclei, the nuclear matter, and the NS properties, out of which only a few EOSs can reproduce the properties consistently with empirical/experimental data \cite{Dutra_2012, Dutra_2014}. To achieve the mass of the secondary component of the GW190814 event, one has to include stiff EOS, which may or may not satisfy different constraints such as flow \cite{Danielewicz_2002}, GW170817 \cite{Abbott_2017, Abbott_2018}, x-ray \cite{Antoniadis_2013, Cromartie_2019}, and NICER data \cite{Miller_2019, Riley_2019}. For this, we search for stiff relativistic mean-field (RMF) EOSs and find that the linear Walecka models (L-RMF) \cite{Serot_1986}, the standard nonlinear $\sigma$-self-coupling  interactions (NL-RMF) \cite{Boguta_1977} and the density-dependent RMF (DD-RMF) \cite{Huang_2020,Rather_2021} sets generally reproduce the maximum  mass of the NS more than 2.50 $M_\odot$. The BigApple force is designed to reproduce the maximum mass of the NS more than 2.50 $M_\odot$ \cite{Fattoyev_2020, DasBig_2020}. Although the L-RMF model predicts larger mass, it fails to reproduce the properties of the finite nuclei and nuclear matter. The recent work of Huang {\it et al.} indicates that the possibility of the secondary object in GW190814 event as pure hadronic NS with DD-RMF parameter sets  \cite{Huang_2020}. In DD-RMF family, the DD-LZ1 and DD-MEX forces reproduce the NS masses as 2.554 and 2.556 $M_\odot$, respectively. Thus, the DD-RMF sets could predict only the lower limit mass of the secondary component in the GW190814 event. Hence, it ruled out the possibility of the hybrid star or any admixture of exotic components such as hyperons, dark matter, etc., inside the NS. Therefore, the L-RMF and DD-RMF forces are not very useful for the investigation.
	%%%%
	\begin{figure*}
		\centering
		\includegraphics[width=0.7\textwidth]{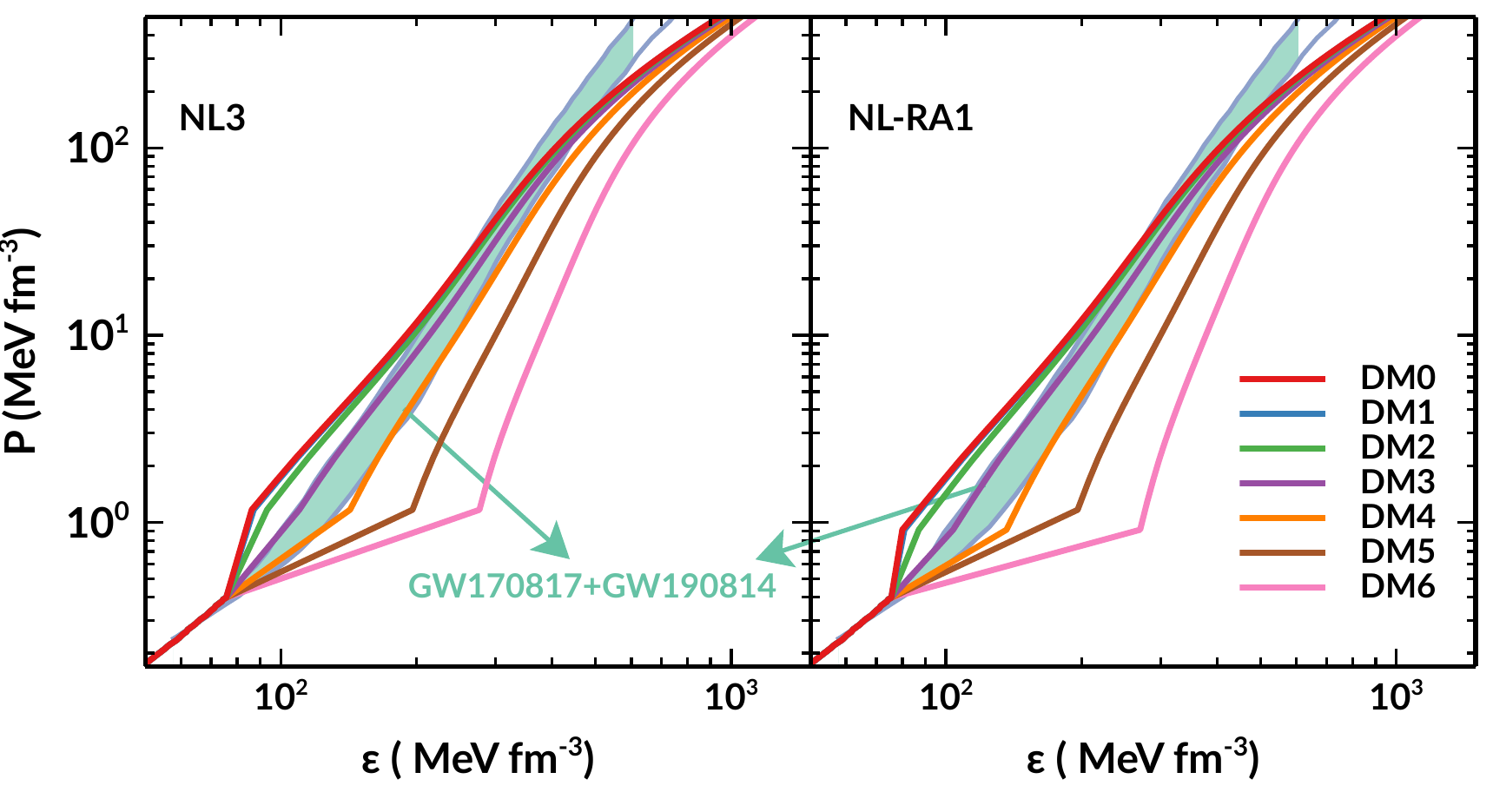}
		\caption{The EOS are shown for NL3 and NL-RA1 with DM Fermi momenta 0--0.06 GeV. The joint constraints from the gravitational wave data GW170817 and GW190814 in shaded region are adopted from Ref. \cite{RAbbott_2020}.}
		\label{fig:eos}
	\end{figure*}
	%%%%
	
	In the present study, we consider the NL-RMF parameter sets as our primary input to achieve the mass of the secondary component. 
	Some of these forces are  NL3 \cite{Lalazissis_1997}, NL3* \cite{Lalazissis_2009}, NL1 \cite{Reinhard_1986}, NL-SH \cite{Sharma_1993}, NL3-II \cite{Lalazissis_1997} and NL-RA1 \cite{Rashdan_2001}, which are generally considered to be stiff EOSs.
	The maximum masses of all these parametrizations  fall in the range of 2.7--2.8 $M_\odot$, which are slightly more massive than the secondary component of the GW190814 event. Therefore, we have two possible scenarios to reduce the maximum mass and radius of the NS (i) with the addition of hyperons or (ii) with the addition of DM. With the addition of exotic particles, the EOS becomes softer, which reduces the mass, radius, and tidal deformability of the NS \cite{Weissenborn_2012, Biswal_2019, Biswalaip_2019, Das_2019, Quddus_2020, Das_2020, das2021effects}. However, in the case of hyperons inside the NS, the predicted canonical radius for NL-RMF type sets found to be $\sim15$ km, which does not lie in the range given by the NICER (11.96--14.26 km) \cite{Miller_2019} and GW190814 (12.2--13.7 km) \cite{RAbbott_2020} predictions. Therefore, in this case, we exclude the hyperons and involve the dark matter as an extra candidate inside the NS.
%%%%
\section{Model}
Dark matter (DM) accounts for $\sim$85\% of the composition in the Universe. When a compact star rotates in the Galaxy through the DM halo, it captures some of the DM due to its enormous gravitational potential and huge baryon density \cite{Goldman_1989, Kouvaris_2011}. We take nonannihilating weakly interacting massive particles (WIMPs) as a DM candidate, because WIMPS are the thermal relics and most abundant DM particle. Different models have already been incorporated to know its actual nature and its effects on the NS properties \cite{Sandin_2009,De_Lavallaz_2010,Kouvaris_2011,Ciarcelluti_2011,Leung_2011,AngLi_2012,Panotopoulos_2017,Ellis_2018,Das_2019, Bhat_2019, Ivanytskyi_2019, Quddus_2020, Das_2020, Das_2021, das2021effects}. We modeled a Lagrangian density by assuming that the DM particles interact with nucleons via standard model (SM) Higgs \cite{Panotopoulos_2017, Das_2019, Das_2020, Das_2021, das2021effects}. 
	%%%%
	\begin{figure}[h]
		\centering
		\includegraphics[width=0.45\textwidth]{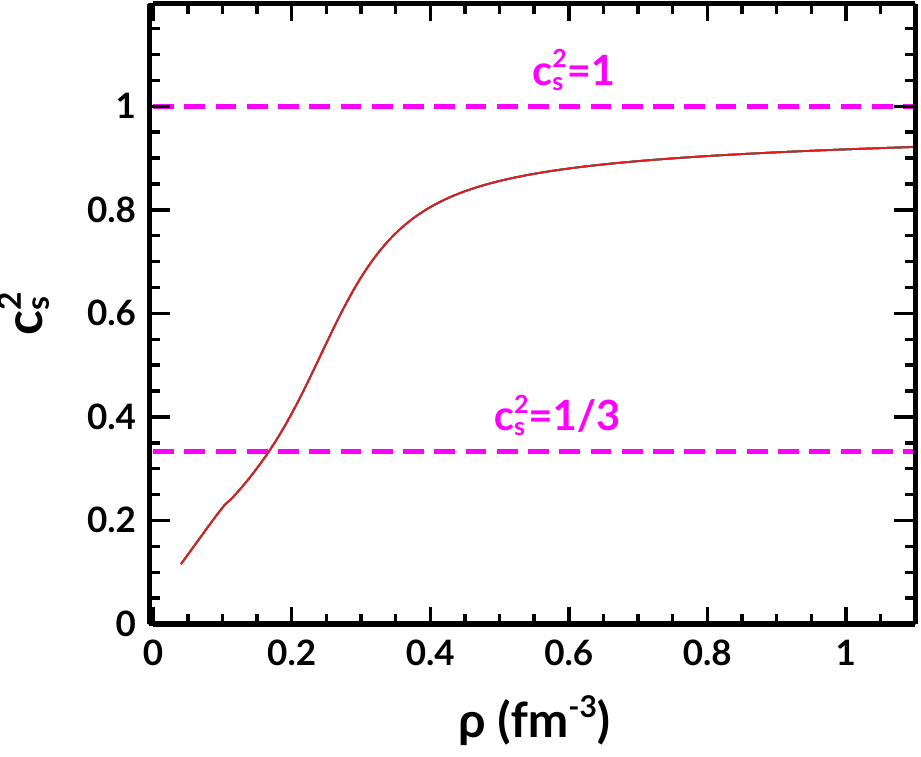}
		\caption{The speed of sound for NL3 parameter set with DM Fermi momenta 0--0.06 GeV. The QCD conformal limit ($c_s^2=1/3$) and maximum speed limit ($c_s^2=1$) are shown with dashed magenta line.}
		\label{fig:causa}
	\end{figure}
	%%%%%%%%%%%%%%
	\begin{figure*}
		\centering
		\includegraphics[width=0.7\textwidth]{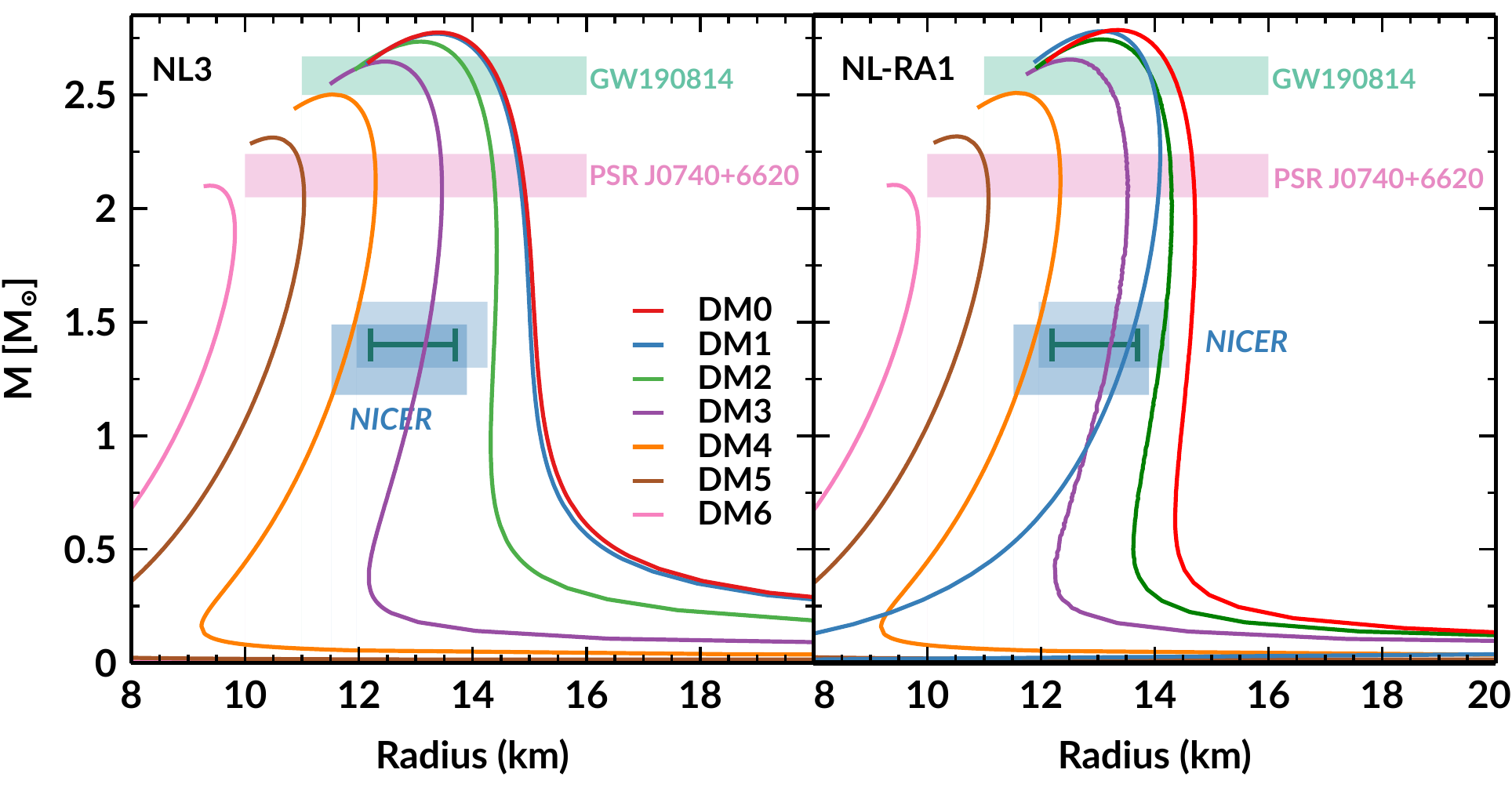}
		\caption{The $M-R$ relations for NL3 and NL-RA1 with different DM Fermi momenta. The horizontal bars represent the maximum mass constraints from the PSR J0740+6620 (light pink) and the GW190814 event (dark cyan). The NICER data are also shown with two boxes from two different analyses \cite{Miller_2019,Riley_2019}. The double-headed green line represents the radius constraints by the GW190814 \cite{RAbbott_2020} for 1.4 $M_\odot$ NS.}
		\label{fig:mr}
	\end{figure*}
   %%%%%
	\begin{eqnarray}
		{\cal{L}}& = & {\cal{L}}_{NM} + \bar \chi \left[ i \gamma^\mu \partial_\mu - M_\chi + y h \right] \chi +  \frac{1}{2}\partial_\mu h \partial^\mu h 
		\nonumber\\
		&&
		- \frac{1}{2} M_h^2 h^2 + f \frac{M_{nucl.}}{v} \bar \varphi h \varphi , 
		\label{eqdm}
	\end{eqnarray}
	where ${\cal{L}}_{NM}$ represents the nuclear matter Lagrangian as given in Refs. \cite{Das_2021, das2021effects}. $\varphi$ and $\chi$ are the nucleon and DM wave function respectively. The mass of the DM ($M_\chi$) is 200 GeV taken in the present calculation. The $y$ and $f$ are the DM-Higgs coupling ($y=0.07$) and nucleon-Higgs form factor ($f=0.35$) respectively. $M_{nucl.}$ is the mass of the nucleon as 939 MeV. The quantities such as $h$, $M_h$, and $v$ are the Higgs field, mass of the Higgs ($M_h=125$ GeV) and Higgs vacuum expectation value ($v=246$ GeV) respectively. We constrained the values of $y$ and $f$ by calculating the scattering cross sections of the DM nucleons which can be written as \cite{Cline_2013}
	\begin{equation}
	    \sigma_{SI}=\frac{y^2f^2M_{nucl.}^2}{4\pi}\frac{\mu_r}{v^2M_h^2},
	\end{equation}
	where $\mu_r$ is the reduced mass. The calculated cross sections for $M_\chi=200$ GeV is found to be $9.70\times10^{-46}$ cm$^{2}$, which is well consistent with XENON-1T \cite{Xenon1T_2016}, PandaX-II \cite{PandaX_2016} and LUX \cite{LUX_2017} with 90\% confidence level. The LHC had also given a limit on the WIMP-nucleon scattering cross section in the range from $10^{-40}$ to $10^{-50}$ cm$^{2}$ \cite{Djouadi_2012}.  From the lattice QCD \cite{Czarnecki_2010} and MILC collaboration results \cite{MILC_2009}, the value of $f$ is constrained to be $f=0.33_{-0.07}^{+0.30}$ \cite{Aad_2015}. The taken value $f=0.35$ in this calculation lies in the region. Thus, we constrain the values of $y$ and $f$ from both direct detection experiments and LHC searches.
	
	Inside the NS, a neutron decays to proton, electron, and antineutrino. The reverse process is also occurring to maintain both $\beta$ equilibrium and charge neutrality conditions. Hence, the total energy density $\cal{E}$ and pressure $P$ for the DM admixed NS involves the contribution from nucleons, leptons, and DM, as given in \cite{Das_2020, Das_2021, das2021effects}.
	%%%%
	\begin{eqnarray}
		{\cal{E}}& = &  {\cal{E}}_{NM} +{\cal{E}}_{l}+\frac{1}{\pi^2}\int_0^{k_f^{DM}} dk \ k^2  \sqrt{k^2 + M_\chi^{\star 2} } 
		 \nonumber \\
		&+& \frac{1}{2}M_h^2 h_0^2 ,
		\label{etot}
	\end{eqnarray}
	and
	\begin{eqnarray}
		P& = &  P_{NM} +P_l+ \frac{1}{3\pi^2}\int_0^{k_f^{DM}} \frac{dk \ k^4} {\sqrt{k^2 + M_\chi^{\star 2}}} 
		\nonumber\\
		&-& \frac{1}{2}M_h^2 h_0^2 ,
		\label{ptot}
	\end{eqnarray} 
	where the ${\cal{E}}_{NM}$ (${\cal{E}}_{l}$) and $P_{NM}$ ($P_l$) are the energy density and pressure for nuclear matter (lepton) \cite{Das_2020, Das_2021, das2021effects}. $k_f^{DM}$ and $M_\chi^*$ are the Fermi momentum and the effective mass of the DM respectively.
	
	We calculate the core EOS for DM admixed NS with NL-RMF family of  parameter sets. For the crust part, we use the Barcelona-Catania-Paris-Madrid (BCPM) equation of state \cite{BKS_2015} and form an unified EOS for the NS. The outer crust is composed of the neutron-rich isotopes embedded in the Coulomb lattice in the degenerate electron gas. As a result, the nuclei capture electrons and form different isotopes in the region $26\leq Z \lesssim 40$. The degenerate electrons solely determine the pressure in this region. Hence, the EOSs for this region are relatively well known \cite{BPS_1971, BKS_2015}. With a density of more than $2.6155\times10^{-4}$ fm$^{-3}$, the nuclei are highly neutron rich, and they do not bind the additional neutrons. Therefore, the neutrons are dripped out from the nuclei. The region defines the boundary between the outer and inner crust. As the densities increase further, the nucleons form clusters in Coulomb crystals of neutron-rich nuclei embedded by uniform electrons. The minimization of the systems' energy produces different exotic structures known as pasta \cite{Ravenhall_1983, Lorenz_1993}. Thus, the detailed knowledge of the EOSs in this region is essential for studying NS properties.
	%%%%%%%
	\section{Results and Discussions}
	%%%%%
	In this section, we calculate the NS properties by using NL-RMF-type EOSs. As it is mentioned, all these sets predict the maximum NS mass more than $2.67 \ M_\odot$. In Fig. \ref{fig:eos}, the EOSs (such as NL3 and NL-RA1) are depicted with $k_f^{DM}$ in the range of 0--0.06 GeV, which are shortly represented as DM0--DM6. The EOSs become softer with the increase of $k_f^{DM}$. The GW190814 discovery provides a joint constraint for NS-black hole (NSBH) merger  \cite{RAbbott_2020}. It is assumed to be a NS only when the maximum mass is not less than the secondary component of GW190814. The EOSs that correspond to DM3 and DM4 almost pass through the joint constraints imposed by GW170817 and GW190814. It indicates that this amount of DM (0.03--0.04 GeV) may be available inside the NS. To test the validity of the EOS, we calculate the speed of sound ($c_s^2=\partial P/\partial \mathcal{E}$) for DM admixed NS. We find that our model respects the causality, which is shown in Fig. \ref{fig:causa}. The value of $c_s^2$ increases up to 0.4 fm$^{-3}$, and it becomes constant beyond that.
	%%%%
	\begin{figure*}
		\centering
		\includegraphics[width=0.7\textwidth]{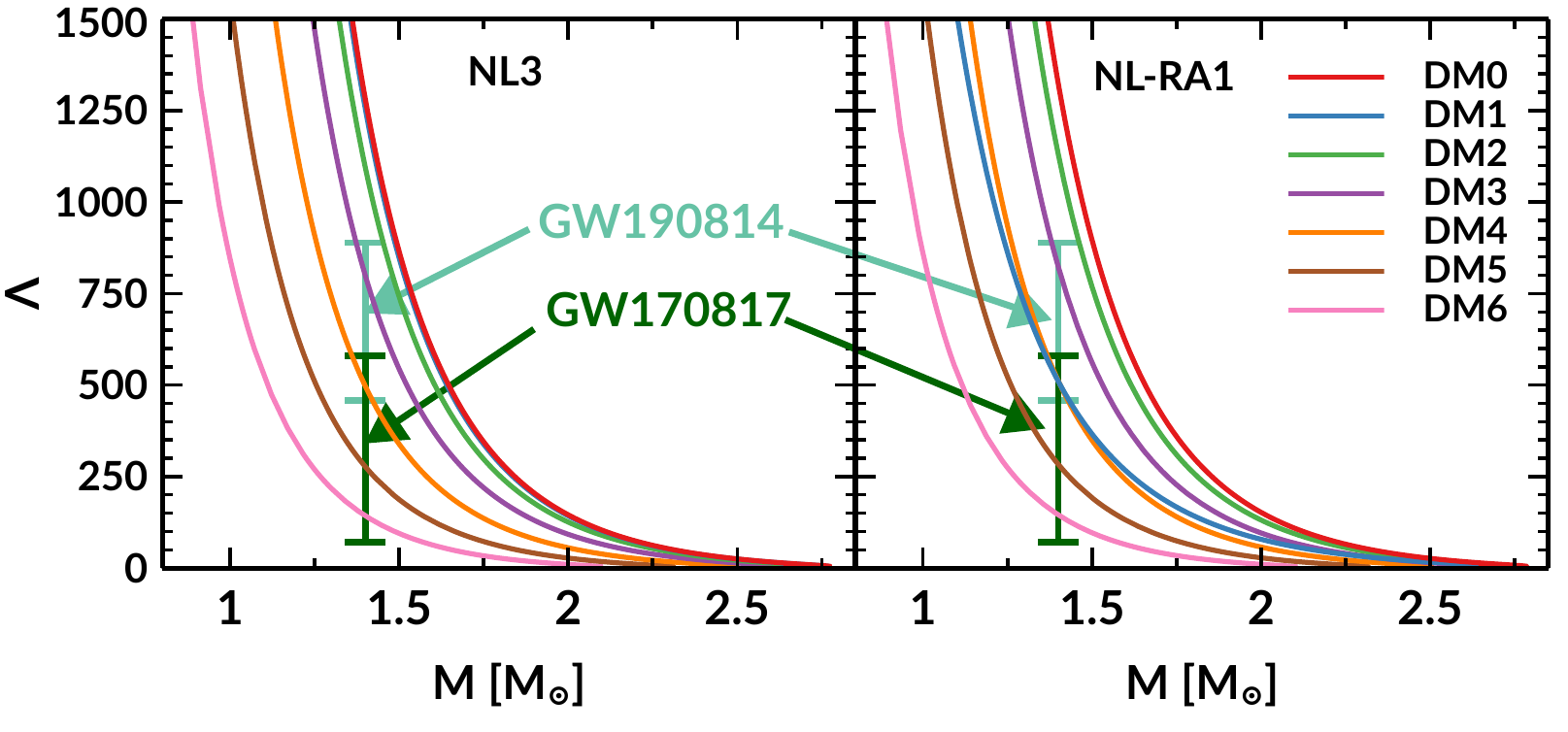}
		\caption{(colour online) The tidal deformability of the NS with the mass for NL3 and NL-RA1 parameter set for different DM Fermi momenta. The dark cyan (green) line implies the constraint given by GW190814 (GW170817) data for 1.4 $M_\odot$ NS.}
		\label{fig:tidal}
	\end{figure*}
	%%%%%%%%%%%%%%
	
	The macroscopic properties such as mass ($M$), radius ($R$), and dimensionless tidal deformability ($\Lambda$) are also calculated with different DM fractions. By solving Tolman-Oppenheimer-Volkoff equations \cite{TOV1, TOV2}; one can find the mass-radius ($M$-$R$) relation for each EOSs by assuming the $P(0)=P_C$ and $P(R)=0$ as the boundary conditions. We find that the NS's mass and radius reduces with DM's addition, as shown in Fig. \ref{fig:mr}. The maximum mass of the secondary object and the canonical radius constraints are 2.50--2.67 $M_\odot$ and 12.20--13.70 km as given in GW190814 \cite{RAbbott_2020}. The $M_{{\rm max}}$ and $R_{1.4}$ range correspond to DM3 are found to be 2.65--2.67 $M_\odot$ and 12.38--12.75 km respectively for the considered parameter sets (see Table \ref{tab:table}), which is consistent with GW190814 data. Under the assumption of a stiff nuclear equation of state, the value of maximum mass with assumed parameter set hints towards the possibility of the secondary component of the GW190814 event to be a dark matter admixed neutron star. The NICER \cite{Miller_2019, Riley_2019} constraints are also respected by DM3 and DM4. Thus we conclude that given a specific model of the nuclear EOS, one can fix the fraction of DM inside the NS from the observational data.
	
	When the NS presents in the tidal field created by its companion star, then it is deformed. The relation between quadruple deformation $Q_{ij}$ and tidal field $\epsilon_{ij}$ is written as \cite{Hinderer_2008, Kumartidal_2017}
	\begin{equation}
	    Q_{ij}=-\lambda \epsilon_{ij},
	\end{equation}
	 where the $\lambda$ is defined as tidal deformability. The dimensionless tidal deformability can be defined as $\Lambda=\lambda/M^5$. We calculate the $\Lambda$ for different DM Fermi momenta as shown in Fig. \ref{fig:tidal}. The values of $\Lambda_{1.4}$ for DM3 are 842.12, 784.59, 932.99, 804.36 and 818.48 for NL3, NL3*, NL-SH, NL3-II and NL-RA1 respectively. The predictions of $\Lambda_{1.4}$ from GW170817 is $190_{-120}^{+390}$ at 90\% confidence level \cite{Abbott_2018}. The GW190814 also put a constraint on $\Lambda_{1.4}$ under NSBH scenario, $\Lambda_{1.4}=616_{-158}^{+273}$ \cite{RAbbott_2020}. The calculated values of $\Lambda_{1.4}$ for DM3 are well matched with GW190814 data, and, for DM4 case, it is well consistent with GW170817 data. This also gives sufficient hints to constraint the amount of DM inside the NS.  
	%%%%
    \section{Summary}
     The mass gap of 2.5--5.0 $M_\odot$ in the secondary component of GW190814 motivates us to explore its true nature, and we find the possibility that the compact object could be a DM admixed NS. To know the mystery of DM inside the NS, we modeled a Lagrangian density by assuming that the DM particles interact with nucleons via SM Higgs. To calculate neutron star EOS, we adopt the RMF model with NL-RMF  parameter sets. The DM admixed NS properties are computed with different DM fractions inside it. 
    
    To achieve the mass of the secondary object, one should take stiff EOSs. In RMF parametrizations, only NL-RMF and DD-RMF sets are stiff in character, which predict the maximum masses more than 2.50 $M_\odot$. Huang {\it et al.} \cite{Huang_2020} also suggested that the possibility of the secondary object as a NS (with DD-RMF sets without any admixture of DM or strange particles). However, the NL-RMF forces with neutron star EOSs predict masses more than 2.67 $M_\odot$ are candidates for DM admixed NS. This scenario is only seen for NL-RMF type sets because no DD-RMF forces predict the mass more than 2.67 $M_\odot$. Hence, we assume that those stars that correspond to DD-RMF sets are only NS without DM inside it.
     %%%%%
	\begin{table*}
	\centering
	\caption{The quantities such as $M_{{\rm max}}$, $R_{{\rm max}}$, $R_{1.4}$ and $\Lambda_{1.4}$ are given with $k_f^{DM}=$ 0, 0.03 and 0.04 GeV for NL3, NL3*, NL1, NL-SH, NL3-II and NL-RA1 respectively.}
	\label{tab:table}
	\renewcommand{\tabcolsep}{0.05cm}
	\renewcommand{\arraystretch}{1.0}
	\begin{tabular}{lllllllllllllllllll}
		\hline\hline
		\multirow{2}{*}{\begin{tabular}[c]{@{}l@{}}$k_f^{DM}$\\   (GeV)\end{tabular}} & \multicolumn{3}{l}{\hspace{1.1cm} NL3} & \multicolumn{3}{l}{\hspace{1.1cm} NL3*} & \multicolumn{3}{l}{\hspace{1.1cm} NL1} & \multicolumn{3}{l}{\hspace{1.1cm} NL-SH} & \multicolumn{3}{l}{\hspace{1.1cm} NL3-II} & \multicolumn{3}{l}{\hspace{1.1cm} NL-RA1} \\ \cline{2-19}
		&0.00 &0.03 &0.04 &0.00 &0.03 &0.04 &0.00 &0.03 &0.04 &0.00 &0.03 &0.04 &0.00 &0.03 &0.04 &0.00 &0.03 &0.04 \\ \hline
		\begin{tabular}[c]{@{}l@{}}$M_{{\rm max}}$\\ $(M_\odot)$\end{tabular} &
		2.78 &2.65 &2.50 &2.76 &2.64 &2.49 &2.84 &2.71 &2.55 &2.79 &2.66 &2.51 &2.77 &2.65 &2.50 &2.78 &2.66 &2.51 \\ \hline
		\begin{tabular}[c]{@{}l@{}}$R_{{\rm max}}$\\   (km)\end{tabular} &
		13.40 &12.46 &11.53 &13.10 &12.38 &11.60 &13.63 &12.75 &11.67 &13.53 &12.63 &11.60 &13.14 &12.42 &11.52 &13.19 &12.47 &11.57 \\ \hline
		\begin{tabular}[c]{@{}l@{}}$R_{1.4}$\\   (km)\end{tabular} &
		14.08 &13.16 &11.82 &14.03 &13.13 &12.13 &14.73 &13.46 &11.90 &14.35 &13.41 &11.90 &14.05 &13.16 &
		11.81 &14.11 &13.22 &11.85 \\ \hline
		$\Lambda_{1.4}$ &
		1311.16 &842.12 &499.04 &1250.45 &784.59 &492.40 &1503.77 &912.65 &548.83 &1528.42 &932.99 &563.90 &1282.90 &804.36 &503.89 &1338.98 &818.48 & 516.96 \\ \hline
	\end{tabular}
    \end{table*}
    %%%%%%

    With the addition of DM, the EOS becomes softer and $M_{{\rm max}}$, $R_{{\rm max}}$, $R_{1.4}$ and $\Lambda_{1.4}$ reduce with increase of DM percentage as given in Table \ref{tab:table}. The EOS corresponds to DM3 and DM4 well passes through the joint constraint inferred from GW170817+GW190814 by assuming that the secondary component is a neutron star only when its maximum mass is not less than the 2.50 $M_\odot$. The causality is also respected by this model for all values of DM Fermi momenta. Other properties such as $M_{{\rm max}}$ and $R_{1.4}$ are well consistent with the GW190814 data for the DM3 case. The values of $R_{1.4}$ for DM3 and DM4 are well constrained with the NICER results in almost all assumed sets. The calculated $\Lambda_{1.4}$ for DM3 and DM4 lie in the range given by both GW190814 and GW170817 data. Thus, one can constrain the DM percentage inside the NS from these observational data.
    
    Therefore, the possibility of the secondary object in GW190814 as a DM admixed NS only when the maximum mass should be more than 2.67 $M_\odot$. We hope future GW and x-ray observations may answer the presence of DM inside the NS and constrain its properties. Thus, we suggest that the LIGO/Virgo have to include DM inside the compact objects when they inferred some properties such as mass, radius, and tidal deformability of the binary neutron star.
   	%\clearpage
	\bibliographystyle{apsrev4-1}
	\bibliography{GW190814}
    \end{document}